\documentclass[pra, showpacs, twocolumn, floatfix,superscriptaddress]{revtex4}
\usepackage{amsmath}
\usepackage{graphicx}
\usepackage{subfigure}
\usepackage{amsmath, amsfonts, amssymb, bm}
\begin{document}
\title{Carrier-envelope phase dependence in single-cycle laser pulse propagation
 with the inclusion of counter-rotating terms}

\author{Ni \surname{Cui}}
\email{ni.cui@mpi-hd.mpg.de}

\author{Mihai A. \surname{Macovei}}
\email{mihai.macovei@mpi-hd.mpg.de}

\affiliation{Max Planck Institute for Nuclear Physics, Saupfercheckweg 1,
D-69117 Heidelberg, Germany}
\date{\today}

\begin{abstract}
We focus on the propagation properties of a single-cycle laser pulse
through a two-level medium by numerically solving the full-wave Maxwell-Bloch equations.
The counter-rotating terms in the spontaneous emission damping are included such
that the equations of motion are slightly different from the conventional Bloch equations.
The counter-rotating terms can considerably suppress the broadening of the pulse envelope and
the decrease of the group velocity rooted from dispersion.
Furthermore, for incident single-cycle pulses with envelope area 4$\pi$,
the time-delay of the generated soliton pulse from the main pulse
depends crucially on the carrier-envelope phase of the incident pulse. This can be
utilized to determine the carrier-envelope phase of the single-cycle laser pulse.
\end{abstract}
\maketitle
\section{Introduction}
The modern technological progress in ultrafast optics makes it possible
to produce few-cycle laser pulses \cite{few-cycle,Atto,THz}.
Recently, a single-cycle pulse with a duration of 4.3 femtoseconds
has been generated experimentally \cite{Single}. Furthermore,
great effort for the generation of extremely short pulses via few-cycle
laser pulses has been made \cite{Kalosha,soliton}, particularly,
single-cycle gap solitons \cite{Macovei} and unipolar half-cycle optical pulses \cite{Song},
respectively, generated in the dense media with a subwavelength structure.
If the pulse duration approaches the optical cycle, the strong-field-matter
interaction enters into the extreme nonlinear optics \cite{eno}, and the
standard approximations of the slowly varying envelope approximation (SVEA),
and the rotating-wave approximation (RWA) are invalid \cite{breakdown,fdtd}.
When the Rabi frequency of the few-cycle laser pulse becomes comparable to
the light frequency, the electric field time-derivative effects will lead to
carrier-wave Rabi flopping (CWRF) \cite{CWRF}, which was observed experimentally
in the semiconductor GaAs sample \cite{GaAs}. In this extreme pumping regime, 
the simple two-level system can still serve as a reference point \cite{eno,ceo,exp}.

For the few-cycle laser pulses, the absolute carrier-envelope phase (CEP) strongly
affects the temporal variation of the electric field. These effects give rise to
many CEP dependent dynamics, such as high-harmonic generation \cite{HHG1,HHG2,HHG3},
optical field ionization \cite{ion1,ion2}, atomic coherence and population transfer
\cite{Wu, Scully}, etc. The CEP dependent strong interactions also provide routines
to determine the CEP of few-cycle ultrashort laser pulses. In particular, the
strong-field photoionization provides very efficient tools to measure the CEP of
powerful few-cycle femtosecond laser pulses for the first time \cite{first}.
Another promising approach to determine the CEP is introduced on the detection of the
THz emission by down-conversion from the few-cycle strong laser pulse \cite{THz-em}.
Recently, the angular distribution of the photons emitted by an ultra-relativistic
accelerated electron also provides a direct way of determining the carrier-envelope
phase of the driving laser field \cite{Piazza}. However, all these measurements of
CEP are based on light amplification in strong-field regime.

Therefore, it is very meaningful to explore routines for determining the CEP
of few-cycle laser pulse at relative lower intensities without light amplification.
The nonperturbative resonant extreme nonlinear optics effects would be good candidates
for measuring the CEP of few-cycle laser pulses with moderate intensities \cite{ceo,exp}.
However, the period of these CEP-dependent effects is $\pi$ due to the inversion symmetry
of light-matter interaction in two-level systems. Thus, the sign of the few-cycle laser
pulse still cannot be determined. In order to remove the $\pi$-shift phase ambiguity,
the violation of inversion symmetry should be considered \cite{metal}. In the presence
of an electrical bias, the phase-dependent signal of ultrafast optical rectification in
a direct-gap semiconductor film implies a possible technique to extract the CEP \cite{Hughes}.
Moreover, the inversion-asymmetry media, such as polar molecules \cite{Yang} and the
asymmetric quantum well \cite{cj}, could also be utilized to determine the CEP of few-cycle
laser pulses.
\begin{figure}[b]
\centering
\includegraphics[width=8.5cm]{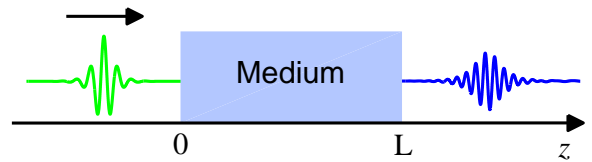}
\caption{\label{fig1} (color online) Schematic of the light-matter interaction scheme.
The green curve illustrates the incident few-cycle laser pulse.
The blue curve illustrates the transmitted laser field.
Symbol $"\rightarrow"$ denotes the propagation direction axis.}
\end{figure}

In this paper, we introduce the counter-rotating terms (CRT) in the spontaneous emission damping,
and investigate the influence of CRT on the propagation dynamics of nonamplified single-cycle laser
pulses in two-level media. The CRT should be considered for such ultrashort pulses interacting with
the medium with strong relaxation processes, because the CRT can notably suppress the broadening of
the pulse envelope and the decrease of the group velocity arising from dispersion. Furthermore, when
the incident single-cycle pulse with envelope area $\Theta=4\pi$ propagates throngh the two-level
medium, it splits into two pulses. The stronger main pulse moves faster than the weaker generated
soliton pulse, and the pulse time-delay between them shows a pronounced CEP dependence. Therefore, in
the presence of a static electric field, we present a simpler approach for measuring the CEP of the
few-cycle laser pulses, by detecting the time-delay of the generated soliton pulse.

\section{Approach}
\subsection{Maxwell Equations}
We consider the propagation of a few-cycle laser pulse in a resonant two-level medium along the $z$ axis,
as shown in Fig.~\ref{fig1}. The pulse initially moves in the free-space region, then it penetrates the
medium on an input interface at $z=0$ and propagates through the medium, and finally, it exits again into
the free space through the output interface at $z=L$. With the constitute relation for the electric
displacement for the linear polarization along the $x$ axis, $D_x=\epsilon_0 E_x+P_x$, the full-wave
Maxwell equations for the medium take the form:
\begin{subequations}
\label{max}
\begin{align}
\frac{\partial H_y}{\partial t}&=-\frac{1}{\mu_0} \frac{\partial E_x}{\partial z},\\
\frac{\partial E_x}{\partial t}&=-\frac{1}{\epsilon_0} \frac{\partial H_y}{\partial z}
-\frac{1}{\epsilon_0} \frac{\partial P_x}{\partial t},
\end{align}
\end{subequations}
where $E_{x}$ and $H_{y}$ are the electric and magnetic fields, respectively. $\mu_{0}$ and $\epsilon_{0}$
are the magnetic permeability and the electric permittivity in the vacuum, respectively.
The macroscopic nonlinear polarization $P_{x} = -Nd_{12}u$ is connected with the off-diagonal
density matrix element $\rho_{12} = \frac{1}{2}(u + iv)$ and the population inversion $w=\rho_{22}-\rho_{11}$,
which are determined by the Bloch equations below.
\subsection{Master Equation}
The Hamiltonian of the two-level system we considered can be described by \cite{kmek}:
\begin{eqnarray}
H&=&\sum_{k}\hbar \omega_{k}a^{\dagger}_{k}a_{k} + \hbar\omega_{0} S_{z}
+ \hbar\Omega(t)(S^{+} + S^{-})\nonumber\\
& + & i\sum_{k}(\vec g_{k}\cdot \vec d_{21})
\bigl \{a^{\dagger}_{k}(S^{+}+S^{-})-{\rm H.c.}\bigr\}, \label{HI}
\end{eqnarray}
where $\omega_{0}$ is the transition frequency, and $\vec d_{21}$ is the electric dipole moment of the
transition between the upper state $|2\rangle$ and the lower state $|1\rangle$. $a^{\dagger}_{k}$
($a_{k}$) is the creation (annihilation) operator for photons with momentum $\hbar k$ and energy
$\hbar \omega_k$, while $\vec{g}_k=\sqrt{\frac{2\pi\hbar \omega_k}{V}}\vec{e}_\lambda$ describes the
vacuum-atom coupling and $\vec{e}_\lambda$ represents the unit polarization vector with
$\lambda\in{\lbrace 1, 2\rbrace}$. $S^{+} = |2\rangle\langle1|$ ($S^{-} = |1\rangle\langle2|$)
is the dipole raising (lowering) operator of the two-level system,
$S_z=(|2\rangle\langle 2|-|1\rangle\langle 1|)/2$ is the inversion operator. $\Omega(t)=d_{12}E_x/\hbar$
is the Rabi frequency of the incident laser field.

In the usual Born-Markov and mean-field approximation, but without the rotating-wave approximation,
the master equation of the system is determined by
\begin{eqnarray}
\dot{\rho}(t)&+&i\left[\omega_0 S_{z} +\Omega(t)\left(S^{+}+S^{-}\right),\rho\right]\nonumber\\
&=& -\gamma \left[S^{+},(S^{+}+S^{-})\rho\right]+{\rm H.c.}, \label{ME}
\end{eqnarray}
where an overdot denotes differentiation with respect to time. Here, $[S^{+},S^{+}\rho(t)]$ and its
hermitian conjugate term represent the counter-rotating terms (CRT) for the spontaneous emission damping,
which are neglected under the rotating-wave approximation when the duration of the laser field pulse $\tau_p$
is much larger than $\omega_{0}^{-1}$. However, for the few-cycle pulses, even the single-cycle or sub-cycle
pulse, the CRT become indispensable and cannot be neglected. In the following, we will investigate the effects
of the CRT on the propagation dynamics of the single-cycle laser pulse in the two-level medium.
\subsection{Bloch Equations}
Based on the master equation (\ref{ME}), including the CRT in the spontaneous emission damping,
the Bloch equations with CRT can be easily derived as follows:
\begin{subequations}
\label{Bloch-CRT}
\begin{align}
\dot{u}&=\omega_0 v,\\
\dot{v}&=-\omega_0 u +2\Omega(t) w-2\gamma_2 v,\\
\dot{w}&=-2\Omega(t) v-\gamma_1 (w+1),
\end{align}
\end{subequations}
where $\gamma_{1}$ and $\gamma_{2}$ are the spontaneous decay rates of the population
and polarization, respectively. The Bloch equations with CRT [Eqs.~(\ref{Bloch-CRT})]
are slightly different from the conventional Bloch equations (see for instance Refs.~\cite{fdtd,Kalosha}):
\begin{subequations}
\label{Bloch}
\begin{align}
\dot{u}&=\omega_0 v-\gamma_2 u,\\
\dot{v}&=-\omega_0 u +2\Omega(t) w-\gamma_2 v,\\
\dot{w}&=-2\Omega(t) v-\gamma_1 (w+1),
\end{align}
\end{subequations}
in which the relaxation constants $\gamma_{1}$ and $\gamma_{2}$ are added phenomenologically.
\begin{figure}[t]
\includegraphics[width=8.0cm]{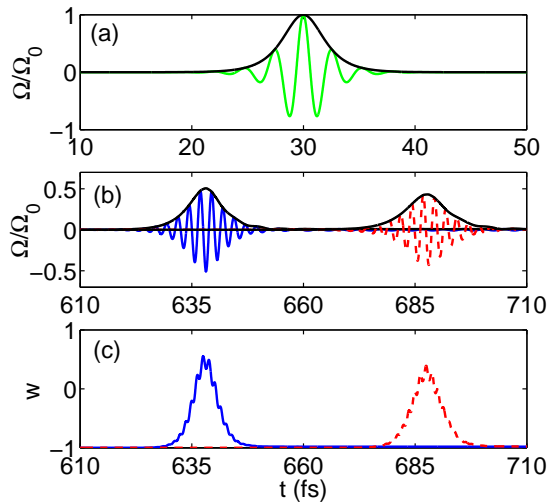}
\caption{\label{fig2} (color online) (a) The Rabi frequency of the incident
single-cycle pulse with envelope area $\Theta=2 \pi$.
(b) and (c) are the time-dependent electric fields and the corresponding
population inversions at the distance $z=90~\rm \mu m$, respectively.
The length of the two-level medium is choosen as $L=110~\rm \mu m$.
The blue solid curves are for the case of Maxwell-Bloch equations with CRT, while
the red dashed curves are for the case with the conventional Maxwell-Bloch equations.
The black lines represent the pulse envelope.}
\end{figure}
\begin{figure}[b]
\centering
\includegraphics[width=7.5cm]{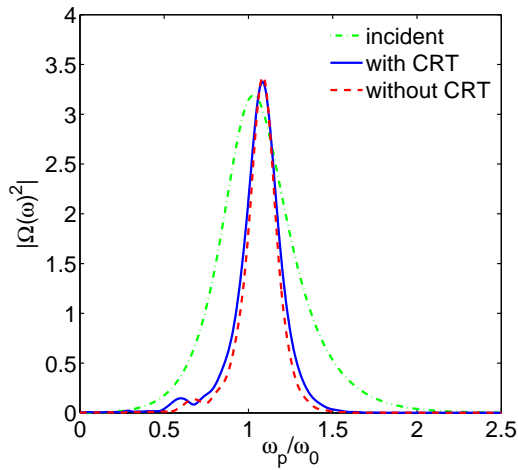}
\caption{\label{fig3} The corresponding spectra of the electric field in Fig.~\ref{fig2}(b).
The black dashed-dotted curve is the spectrum of the incident laser pulse.
The blue curve depicts the case with CRT, and the red dashed curve the case without CRT. }
\end{figure}
\subsection{Numerical Method}
The propagation properties of the few-cycle laser pulse in the two-level medium
can be modeled by the full-wave Maxwell-Bloch equations beyond the SVEA and RWA,
which can be solved by the iterative predictor-corrector finite-difference time-domain
discretization scheme \cite{fdtd,fdtd-method}. For such an extremely short laser pulse,
we define the vector potential at $z=0$ as in Refs.~\cite{liu, niu}:
\begin{eqnarray}
A_x(t)=A_0~{\rm sech} [1.76(t-t_0)/\tau_p]\sin{[\omega_p(t-t_0)+\phi]}, \label{vec}
\end{eqnarray}
where $A_{0}$ is the peak amplitude of the vector potential, $\omega_{p}$ is the photon energy,
and $\phi$ being the CEP. $\tau_{p}$ is the full width at half maximum (FWHM) of the short pulse
and $t_{0}$ is the delay. The electric field can be obtained from $E_x =-\partial A_x(t)/\partial t$.
In what follows, we assume that the two-level medium is initialized in the ground state with $u=v=0$
and $w=-1$. The material parameters are chosen as in Ref.~\cite{Kalosha}:
$\omega_0=2.3~\rm fs^{-1}$ ($\lambda=830~\rm nm$), $d_{12}=2~\times~10^{-29}~\rm Asm$,
$\gamma_1^{-1}=1~\rm ps$, $\gamma_2^{-1}=0.5~\rm ps$, the density $N=4.4~\times~10^{20}~\rm cm^{-3}$.
The incident pulse has a FWHM in single optical cycle $\tau_p=2.8~\rm fs$ and the photon energy
$\omega_p=\omega_0$. The Rabi frequency $\Omega_0=-A_0 \omega_p d/\hbar=1~\rm fs^{-1}$ corresponds
to the electric field of $E_x=5~\times~10^9~\rm V/m$ or an intensity of $I=6.6~\times~10^{12}~\rm W/cm^2$,
and the incident pulse area is defined as $\Theta=\int^{\infty}_{-\infty}\Omega(t)dt$.

\section{Results and discussion}
Now, we focus on the effects of CRT on the propagation dynamics of single-cycle laser pulses
in two-level medium by comparing the numerical results from the Maxwell-Bloch equations
with CRT [Eqs.~(\ref{max}) and (\ref{Bloch-CRT})] and without CRT [Eqs.~(\ref{max}) and (\ref{Bloch})].
We use an incident single-cycle pulse with envelope area $\Theta=2\pi$ for these simulations with the
medium zone length: $z=110~\rm \mu m$.

According to the standard area theorem, the pulse with area $\Theta=2\pi$ can transparently propagate
through the two-level medium without suffering significant lossness - the so-called self-induced
transparency (SIT) \cite{sit}. However, when the laser pulse envelope contains only few optical cycles,
the standard area theorem breaks down because of the occurrence of CWRF \cite{CWRF}. From our numerical
results, for the short propagation distance, the usual SIT regime is essentially recovered. However, at
a further distance, the established conditions for SIT are destroyed due to the extreme nonlinear optical
effects. Fig.~\ref{fig2}(b) and Fig.~\ref{fig2}(c) present the normalized electric-field pulses and
the corresponding population inversions at the distance $z=90~\rm \mu m$ for different approaches, namely,
the blue solid curves depict the case obtained from Maxwell-Bloch equations with CRT, while the red dashed
curves are for the conventional approach without CRT. Compared with the incident single-cycle $2\pi$ pulse
in Fig.~\ref{fig2}(a), the electric-field pulses for both two cases become clearly broadened induced by the
dispersion, and suffer the decrease in pulse amplitude. Accordingly, the population differences for both
cases undergo an incomplete Rabi flopping with the CWRF [Fig.~\ref{fig2}(c)].
\begin{figure}[t]
\centering
\includegraphics[width=8.5cm]{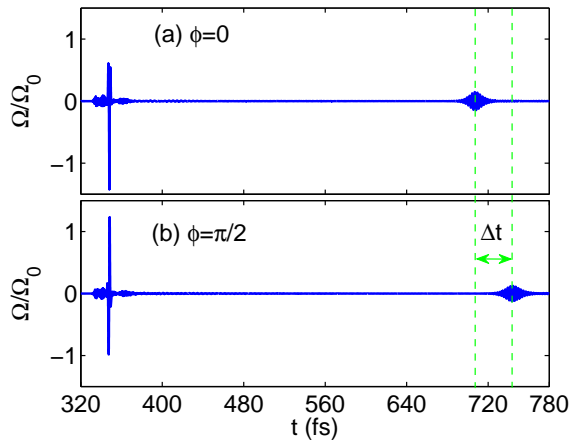}
\caption{\label{fig4} The transmitted electric field of the incident pulses with envelope area
$\Theta=4\pi$ for different initial CEPs: (a) $0$ and (b) $\pi/2$. The length of two-level medium is
$L=80~\rm \mu m$.}
\end{figure}

However, there are notably different features between these two approaches.
The electric-field pulse from the approach with CRT [blue solid curves in Fig.~\ref{fig2}(b)]
is evidently narrower than that in the case without CRT [red dashed curves in Fig.~\ref{fig2}(b)].
This can be easily seen from the corresponding spectra shown in Fig.~\ref{fig3}.
The spectrum of the case with CRT is obviously more broadened than that in the case without CRT,
although both of them become narrower than the incident spectrum.
The envelope peak from the approach with CRT is relatively larger
than that in the case without CRT [Fig.~\ref{fig2}(b)],
hence, the former case lead to more population inversion
at the leading edge of the electric-field pulse [Fig.~\ref{fig2}(c)].
Moreover, there is a notably time delay of the electric-field pulses and the
corresponding population inversions between the two approaches [Fig.~\ref{fig2}(b) and (c)].
It means that the group velocity of the propagating pulse from the conventional approach
without CRT is obviously smaller than that in the case with CRT. This difference in the group velocity
is rooted from the different influence of dispersion effects for these two approaches.
Comparing Eqs.~(\ref{Bloch-CRT}) with Eqs.~(\ref{Bloch}), there is no damping of the real part of
polarization $u$ in the Bloch equations with CRT, which indicates that the dispersion does not
suffer lossness. That is to say, the presence of CRT evidently suppresses the strong dispersion effects,
which lead to the broadening of pulse envelope and the decrease of the group velocity.

In addition, we also find that the influence of CRT on the propagation dynamics
of the single-cycle laser pulses is significantly enhanced with the increase of the spontaneous
decay rates. Therefore, the CRT is important and indispensable for the study of the propagation
properties of few-cycle laser pulses in the medium with strong relaxation processes.
In the following discussion, we will use our established full-wave Maxwell-Bloch equations with CRT
[Eqs.~(\ref{max}) and (\ref{Bloch-CRT})] to explore an approach for determining the CEP of the
single-cycle laser pulse.

In what follows, we simulate the incident single-cycle pulses with larger envelope area, i.e.
$\Theta=4\pi$, propagating through the two-level medium with a length $L=80~\rm \mu m$.
During the course of pulse propagation, the medium absorbs and emits photons and redistributes
energy in the pulse. The propagating pulses are altered in shape until it reaches a stable status
after some propagation distance by splitting into two pulses, the strong main pulse and the SIT
soliton pulse. However, the former moves faster than the latter, which is why the generated SIT soliton
pulse breaks up from the main pulse. We show the transmitted pulses of the incident single-cycle pulse
with pulse envelope area $\Theta=4\pi$ for different CEP $\phi=0$ and $\phi=\pi/2$ in Fig.~\ref{fig4}.
It can be seen that both of the transmitted pulses split into two pulses. There is a time delay between
the main pulse and the soliton pulse defined as $t(\phi)$. The time-delay for the incident pulse with
CEP $\phi=\pi/2$ [$t(\phi=\pi/2)$] is evidently larger than that in the case with CEP $\phi=0$ [$t(\phi=0)$].
It demonstrates that the pulse time delay $t(\phi)$ is sensitive to the initial CEP of the incident pulse.

\begin{figure}[b]
\centering
\includegraphics[width=8.0cm]{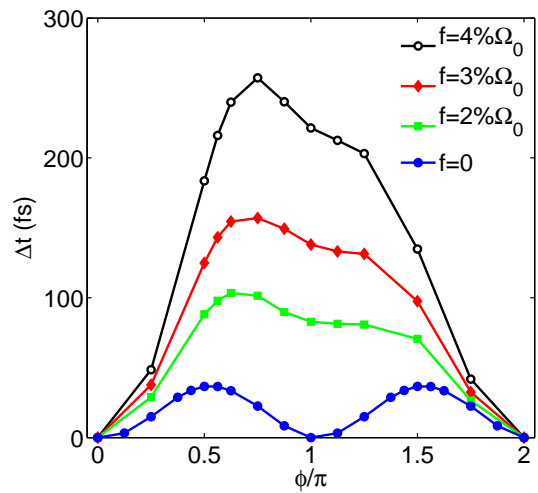}
\caption{\label{fig5} The relative pulse delay between the transmitted soliton pulses
as a function of the CEP of the incident pulses for different strengths of the static electric field.
The solid lines are a guide for the eyes.}
\end{figure}

For simplicity, we define the relative pulse time delay $\Delta t=t(\phi)-t(\phi=0)$ to indicate
the CEP dependence. We present the relative pulse time delay $\Delta t$ as a function of the
initial CEP of the incident pulse in Fig.~\ref{fig5} with blue circles. It is found that the relative pulse
time delay $\Delta t$ is related to the CEP of the incident pulse with a nearly cosinelike dependence.
However, the time delay $t(\phi=\pi)$ is exactly the same as $t(\phi=0)$, and hence, the period of the
CEP-dependent pulse time delay is only $\pi$ because of the inversion symmetry of light-matter interaction.
This means that we cannot distinguish the incident pulse from the initial CEP $\phi~\rightarrow~\phi+\pi$.

In order to remove the $\pi$-shift phase ambiguity, we add a static electric field to break the inversion
symmetry of the light-matter interaction \cite{static}. As a result, the Rabi frequency terms in Bloch
equations [Eqs.~(\ref{Bloch-CRT})] should change as $\Omega(t)~\rightarrow~\Omega(t)+f$, where $f$ describes
the strength of the static electric field. The presence of the static electric field gives rise to the
enhancement of the CEP-dependent variation in the peak electric strength of the single-cycle pulse, which
will enhance the CEP dependence of the dynamics effects. The relative pulse time delay $\Delta t$ of the
transmitted soliton pulses as a function of the initial CEP of the incident single-cycle pulses for
different static electric fields is presented in Fig.~\ref{fig5}. Compared with the blue circles of $f=0$,
the influence of the static electric field on the relative pulse time delay is significant. Let us take
$f=2\%~\Omega_0$, for example green squares in Fig.~\ref{fig5}, then the relative time delay $\Delta t\neq 0$
at $\phi=\pi$, i.e., $t(\phi=\pi)$ is quite different from $t(\phi=0)$ in the presence of the static
electric field. The variation with the CEP of the incident pusle also becomes much stronger. The period of
the CEP-dependent pulse time delay becomes $2\pi$ because the inversion symmetry is broken assisted by
the static electric field. Moreover, with the increase of the static electric field, such as $f=3\%~\Omega_0$
and $f=4\%~\Omega_0$, the dependence of the relative pulse time delay on the initial CEP is further enhanced
[red diamond and black circles in Fig.~\ref{fig5}].

As a result, in the presence  of the static electric field, if the relative time delay of the generated
soliton pulses is calibrated, this effect suggests an approach for determining the CEP of the incident
single-cycle laser pulses in both sign and amplitude. In addition, it should be pointed out that the
pulse time delay might be much easier to detect compared with other features of the soliton pulse, such as
the intensity and pulse duration \cite{Yang}. Finally, in our discussion, the static electric filed strengths, 
which are a few percentages of the single cycle laser pulse strength, exceed a few MV/cm. In order to achieve 
this kind of strength of the static electric field in an experiment, we may proceed with a special case as 
suggested in Ref.~\cite{static} where an additional electric field with a much lower frequency 
(such as $CO_2$ laser field, terahertz pulses or a midinfrared optical parameter amplifier pulse) 
is used instead of the static electric field. The ultra-short dynamics can prevent the system from being 
destroyed or ionized.

\section{Summary}
In summary, we investigated the propagation properties of single-cycle laser pulses
in a two-level medium including the counter-rotating terms in the spontaneous emission
damping. We found that the counter-rotating term can efficiently suppress the
broadening of the pulse envelope and the decrease of the group velocity. Thus, the
counter-rotating term is important and indispensable for the study of the propagation
dynamics of few-cycle laser pulses, even for single-cycle and sub-cycle pulses.
Furthermore, we explored the CEP-dependence of the generated soliton pulse from the
single-cycle laser pulse propagating through the two-level medium. The time delay of
generated soliton pulses depends sensitively on the CEP of single-cycle incident laser
pulse. Hence, the presence of the static electric field enhances the CEP-dependence of
the relative pulse time delay, which have an potential application in determining the CEP
of the incident single-cycle laser pulse.


\end{document}